# Neural Network Tracking of Moving Objects with Unknown Equations of Motion

Boaz Fish   and   Ben Zion Bobrovsky

(boazfish@mail.tau.ac.il)   (bobrov@eng.tau.ac.il)

Faculty of Engineering, Tel Aviv University

February 2020

ABSTRACT

In this paper, we present a Neural Network design that can be used to track the location of a moving object within a given range based on the object's noisy coordinates measurement. This function is commonly performed by the Kalman filter. Our goal in this paper was to present a network that can outperform the Kalman filter in this function for certain scenarios. From our knowledge in control theory, we know that for a test path that has known equations of motion, the Kalman filter will always produce the best results. If the path and the measurement are linear, then the classic Kalman filter can be used. If the path is nonlinear, a variation of the Kalman filter can be used, most commonly the Extended Kalman Filter (EKF). However, for other test paths that have unknown equations of motion, it is possible to get more accurate results than the Kalman filter. We present in this paper results showing that for a number of test paths that have unknown equations of motion, our Neural Network provides a better location estimation than the ideal Kalman filter. We also compare our results to a number of other tracking methods, all of which are also outperformed by our Neural Network. When comparing the root mean square MSE measurement, we see that our network outperforms the Kalman filter by up to 60% for certain paths. Our network uses prior knowledge on the location range of the path as well as possible patterns that may show up in the path to correctly identify the location of the moving object. This gives our method an advantage over the Kalman filter, which does not use any prior knowledge regarding the location range of the moving object, or of possible patterns that may appear in the path. Similarly, the other tracking methods that we used for comparison also do not use any of this prior knowledge in order to perform the tracking. It is for this reason that our network is able to outperform these other methods when testing them with paths with unknown equations of motion.

MOTIVATION

Throughout our research, our main goal was to develop a method to accurately track the location of a moving object using noisy measurements, when the equations of motion are unknown. When the equations of motion are known, there is a known solution to this problem, as the Kalman filter can be used. For cases when the equations of motion are unknown, we are interested in a different method. While the Kalman filter can be used to solve this problem as well, we believe that there is a more efficient way of solving it. Unlike the cases where the equations of motion are known, there is no guarantee that the Kalman filter produces the best results in these cases. Our goal is to find a method that outperforms the Kalman filter in these cases, in order to find the best estimation. Additionally, we want this method to perform reasonably well for cases when the equations of motion are known, despite never being able to reach the accuracy of the Kalman filter.

RELATED WORK

In the past, there have been researches conducted where Neural Networks have been trained to attempt to replace the Kalman filter. Our research is different from previous researches in terms of architecture, training methods, and the exact way it is used. Most of the references for these projects did not present full results, so it is hard to say exactly how successful they are. However, our results look promising compared to the results that we have seen.

In [1] and [2] the architecture of the networks is similar to our network. However, the training methods used are different from ours, which leads to different results. The results in these experiments are terrific for certain test paths (that are comprised mostly of straight lines), though not as well for other types of paths.

In [3] and [4] we can see similar experiments to our research, however with different styles of networks, varying in sizes. The training methods and results are unknown.

INTRODUCTION

In this paper, we built a Neural Network that is designed to estimate the location of a moving object based on noisy coordinate measurements. An example for this problem in real life can be seen when trying to track the exact location of a moving vehicle within a given range. The most common method used for this purpose in many cases similar to ours, is the Kalman filter. This method is used for the estimation of the location of the moving object, and it often produces excellent results. When the movement of the object can be represented by known equations of motion, the Kalman filter and its derivatives produce ideal results. However, when the moving object has unknown equations of motion, the Kalman filter becomes less effective. Our goal is to find a more effective way to estimate the location in such cases using Neural Networks.

When a moving object is moving according to given equations of motion, it becomes easier to accurately predict the rest of the path of the moving object. The equations of motion give us additional information about the moving object. Even if we are measuring the location of the object with added noise, the fact that we have additional information about the expected movement makes it easier to track. This is exactly why tracking objects with unknown equations of motion is so difficult. When the only information we have is a noisy measurement of the current location, the estimation of the exact result is less than ideal. This is the reason that it is necessary to find a method that works for moving objects that don't have known equations of motion.

In our research, we simulate this problem and try to solve it using Neural Networks. We simulate these noisy measurements by adding white Gaussian noise to the ground truth coordinates that we are trying to measure. The size of the Gaussian noise in our simulation is relative to the range that the moving object can be found in. Our main goal is to properly track the location of a moving vehicle travelling at a speed of $10\frac{m}{s}$ within a given range that size of 100m x 100m. For this size range, the Gaussian noise added to the measurements has a magnitude of $\sigma = 1m$ or $\sigma = 2m$ value.

A moving vehicle can move quite randomly in a given range. While we know the speed of the vehicle, we do not know what direction it will go in an any given moment, making its movement random. Due to this randomness, we are unable to write exact equations of motion for the movement of the vehicle. The fact that there are no exact known equations of motion for this process means that the Kalman filter is not guaranteed to be the most effective method to track the location of the vehicle. Our Neural Network is meant to give more accurate measurements that the Kalman filter in this case.

EXPERIMENT

Our experiment is comprised of 3 parts:

1. Training our network.

2. Testing our network with paths that have unknown equations of motion – simplified car model.

3. Testing our network with paths that have known equations of motion.

1. In this part we trained our Neural Network to be able to convert noisy measurements in to x,y coordinates of the moving object. The training is done with a training path designed to prepare the network for any kind of test path. The path is based on small diagonal movements, which is meant to train the network for all paths, as most paths can be built by combining infinitesimal diagonal paths.

The inputs to the network are the noisy measurements, and the x,y coordinates of the 2 previous measurements. The network has 3 layers – an input layer, an output layer, and one hidden layer. The hidden layer has 400 neurons, with the sizes of the input and output layers being set according to the number of inputs and outputs, respectfully. The network was trained for 15,000 iterations

2. In this part we tested our network, as well as other methods, on numerous test paths that have unknown equations of motion – the simplified car model. The methods used were our network, the Kalman filter, and a generic sliding window filter.

We used 2 test paths for this section. The first path, referred to as path #1, is comprised of multiple circular paths. The second path, referred to as test path #2, is comprised of a circular path and diagonal lines. In the following graphs we can see the each of the test paths:

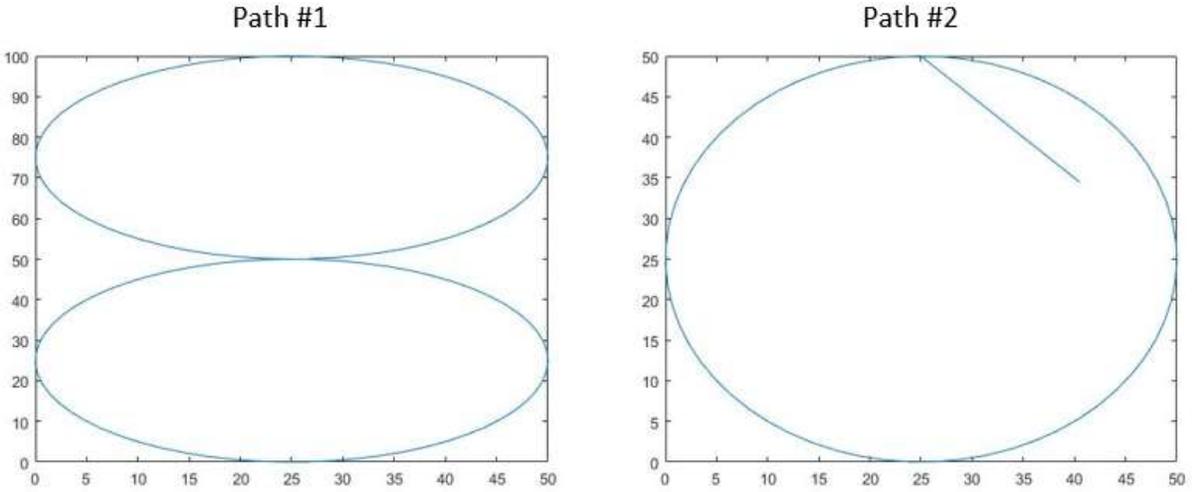

Figure 1: Test paths used – simulated car model

Each of these paths were used for the experiment, with different sized time steps (dt) as well as different multipliers of the measurement noise added to the measurements ($\sigma$).

3. In this part we tested our network, as well as other methods, on numerous test paths that, unlike in part 2, have known equations of motion. The methods used were our network, the Kalman filter, and a generic sliding window filter.

The test paths used for this section were created as follows:

- The path for each axis is represented by the continuous model:
    - $\dot{x}(t) = -a \cdot x(t) + b \cdot n(t)$
    - $y = x + \sqrt{R_c} \cdot v(t)$
- And using discretization the path is represented by:
    - $x[k+1] = (1 - a \cdot dt)x[k] + \sqrt{dt} \cdot b \cdot n[k]$
    - $y[k] = x[k] + \sqrt{\frac{R_c}{dt}} \cdot v[k]$
    - $\sigma = \sqrt{\frac{R_c}{dt}}$ is the multiplier of the measurement's noise.
- These formulas are for each axis, separately.

    We simulated many versions of this experiment, with many different values of each parameter.

RESULTS

The results for the experiments with the simulated car model, with unknown equations of motion, are as follows:

| dt | Path # | $\sigma = \sqrt{\frac{R_c}{dt}}$ | Neural Network RMS | Kalman RMS | Window Filter RMS | RMS Ratio (NN/Kalman) |
|---|---|---|---|---|---|---|
| 0.1 | 1 | 1 | 0.43 | 0.51 | 0.63 | 0.87 |
| 0.1 | 2 | 1 | 0.43 | 0.49 | 0.61 | 0.88 |
| 0.1 | 1 | 2 | 0.70 | 0.83 | 0.91 | 0.85 |
| 0.1 | 2 | 2 | 0.61 | 0.77 | 0.88 | 0.78 |
| 1 | 1 | 1 | 0.65 | 1.53 | 0.89 | 0.43 |
| 1 | 2 | 1 | 0.46 | 1.57 | 0.93 | 0.31 |
| 1 | 1 | 2 | 0.73 | 2.44 | 1.2 | 0.30 |
| 1 | 2 | 2 | 0.66 | 2.31 | 1.17 | 0.29 |

Table 1: results for simulated car model

The visual results for some of the experiments can be seen in the following 2 figures:

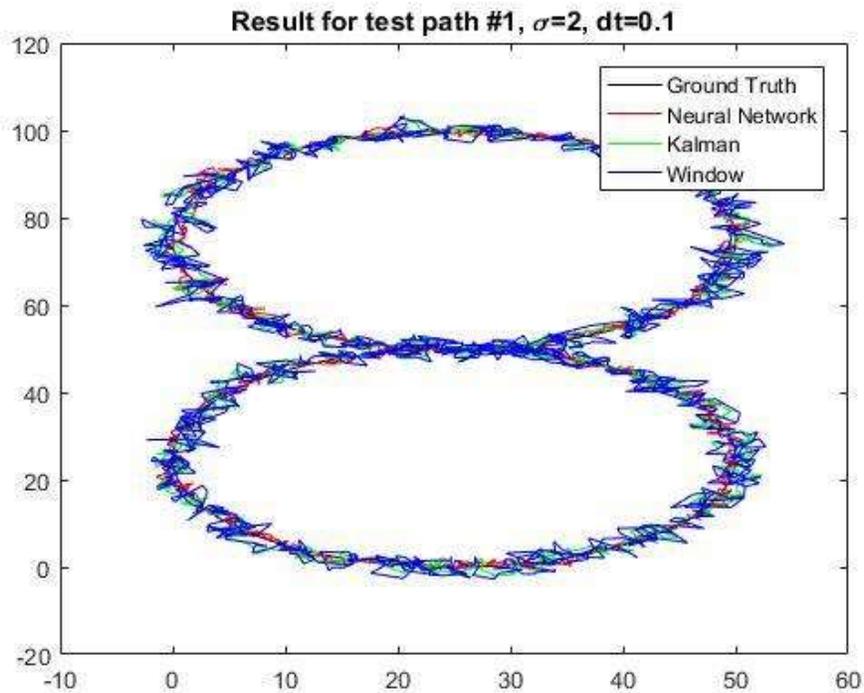

Figure 2: Results for test path #1

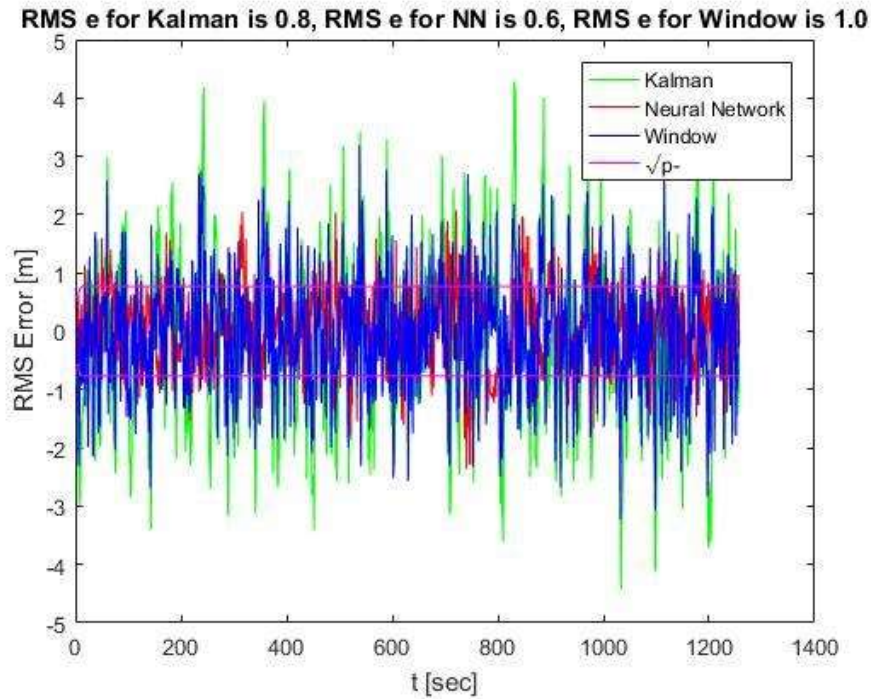

Figure 3: Results for test path #2

These results show us that for all of our test paths, our Neural Network outperforms both the Kalman filter and the Window filter. We can see this is the case, as the RMS error for the Neural Network is lower than all other methods in each experiment. We can also see that the RMS ratio is below 1 for each case, meaning that the Neural Network has a lower RMS error than the Kalman filter. Since our research goal is to find a method that works better for paths with unknown equations of motion, we believe that these experiments show that it is successful.

The results for the experiments with the known equations of motion are as follows:

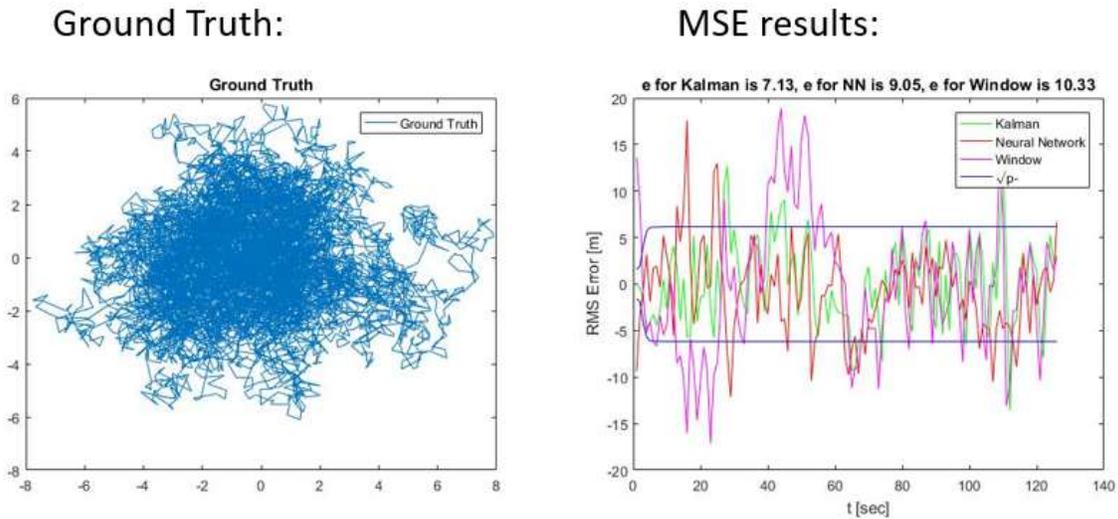

Figure 4: Results for test path with known equations of motion

These results were quite positive as well. On average we got a 1.3 ratio between the RMS error of the Neural Network and the Kalman filter. In all of the experiments, the Kalman filter performed best (as expected), followed by our network, followed by the window filter. Thus, our proposed Neural Network filter is quite close to the optimal (Kalman) filter.

CONCLUSION

In this paper, we created a Neural Network based method for estimating the location of a moving object without known equations of motion. Until now, the Kalman filter and its derivatives have been the primary methods used for this task, as well as cases when there are known equations of motion. When there are known equations of motion, we are guaranteed to get the most accurate results from the Kalman filter. However, when there are no known equations of motion, there is no such guarantee, and it is possible to outperform the Kalman filter.

We presented the results of our Neural Network and compared them to the results of the Kalman filter, as well as the results of a simple sliding window filter. We experimented with both test paths that have known equations of motion, as well as test paths that have no known equations of motion.

For test paths that have known equations of motion, the most accurate results came from the Kalman filter, with our Neural Network still producing results better than the sliding window filter. It was expected that the Kalman filter would produce the best results, though it was still a promising sign that our Neural Network outperformed the sliding window filter.

For test paths that have no known equations of motion, the most accurate results came from our Neural Network, followed by the Kalman filter and the sliding window, producing worse results. We consider this outcome to be a great success, as our goal was to outperform the Kalman filter with our Neural Network in certain scenarios. In our experiments, we can clearly see that our Neural Network outperforms the Kalman filter with all of our test paths. With these results, we believe that we have achieved our goal.

Our network was tested on a limited number of paths. The results are extremely positive, though at this point we cannot yet definitively say that our method will outperform the Kalman filter for all paths with no equations of motion. We believe that in future research we will be able to increase the effectiveness of our Neural Network and show that it is effective for more scenarios without known equations of motion.

The performance of our network leads us to believe that this technology may indeed be superior to all other existing tracking methods. Our method outperforms all methods when there are no equations of motion for our test paths. While this in itself is very promising, it is equally if not more promising that our network performs at a high level for Gaussian paths with known equations of motion. While the results will always be worse than the Kalman filter, the fact that our network is close to the Kalman filter is performance is very promising and gives us much confidence in our results.